\documentclass[english,english,aip,jcp,dvipsnames,preprint]{revtex4-1}
\usepackage[T1]{fontenc}
\usepackage[latin9]{inputenc}
\usepackage{geometry}
\usepackage{color}
\usepackage{float}
\usepackage{amsmath}
\usepackage{amssymb}
\usepackage{graphicx}
\usepackage{subscript}

\makeatletter

\newcommand{\lyxmathsym}[1]{\ifmmode\begingroup\def\b@ld{bold}
  \text{\ifx\math@version\b@ld\bfseries\fi#1}\endgroup\else#1\fi}

\providecommand{\tabularnewline}{\\}

\usepackage{braket}
\makeatother

\usepackage{babel}
\begin{document}

\title{Anharmonic Vibrational Eigenfunctions and Infrared Spectra from Semiclassical
Molecular Dynamics }

\author{Marco \surname{Micciarelli}}
\email{marco.micciarelli@unimi.it}

\author{Riccardo \surname{Conte}}

\author{Jaime \surname{Suarez}}

\author{Michele \surname{Ceotto}}
\email{michele.ceotto@unimi.it}

\affiliation{Dipartimento di Chimica, Università degli Studi di Milano, via C. Golgi
19, 20133 Milano, Italy}
\begin{abstract}
We describe a new approach based on semiclassical molecular dynamics
that allows to simulate infrared absorption or emission spectra of
molecular systems with inclusion of anharmonic intensities. This is
achieved from semiclassical power spectra by computing first the vibrational
eigenfunctions as a linear combination of harmonic states, and then
the oscillator strengths associated to the vibrational transitions.
We test the approach against a 1D Morse potential and apply it to
the water molecule with results in excellent agreement with discrete
variable representation quantum benchmarks. The method does not require
any grid calculations and it is directly extendable to high dimensional
systems. \textcolor{black}{The usual exponential scaling of the basis
set size with the dimensionality of the system can be avoided by means
of an appropriate truncation scheme.} Furthermore, the approach has
the advantage to provide IR spectra beyond the harmonic approximation
without losing the possibility of an intuitive assignment of absorption
peaks in terms of normal modes of vibration.
\end{abstract}
\maketitle

\section{Introduction\label{sec:Introduction}}

Experimental vibrational spectroscopy is a powerful tool used for
countless applications in chemistry and material science. It is routinely
complemented by computer simulations which allow to rationalize the
absorption peaks observed experimentally.\citep{Bowman_Meyer_Polyatomic_2008,Barone_Panek_VPT2Review_2011}
The information extracted from these simulations often consists in
the assignment of the experimental peaks in terms of vibrations of
specific functional groups, especially when dealing with many degrees
of freedom as in the case of materials and large molecular systems.
Many calculations though provide only harmonic estimates of vibrational
energies and motions. However, the harmonic approximation fails in
describing high energy vibrations adequately and, in order to fit
computational data to experiments, harmonic frequencies have to be
often scaled with \emph{ad hoc} procedures\citep{scott_radom_scalingoringinal_1996,Irikura_Kacker_ScalingFactors_2005}
reducing the level of reliability of the calculations. 

In the last decades considerable efforts have been made to develop
classical\citep{Gaigeot_Vuilleumier_IRspectroscopy_2007} and quantum
theoretical methods able to go beyond the harmonic approximation.
Vibrational configuration interaction,\citep{Bowman_Huang_MULTIMODE_2003,carter_sharma_multimode_2012,Yagi_Hirata_VCI/VCC_2012,Thomsen_Christiansen_VCI/VCC_2014}
multi configuration time dependent Hartree,\citep{Meyer_Cederbaum_MCTDH_1990,Meyer_Worth_HighDimMCTDH_2003},
collocation methods,\citep{Avila_Carrington_SmolyakCollocation_2017,Manzhos_Carrington_RectangularCollocation_2016,kosloff_1994}
perturbative approaches like the second order vibrational perturbation
theory (VPT2),\citep{barone_Bloino_IRmediumsized_2014,Barone_Puzzarini_neverendinggly_2013},
path integral molecular dynamics,\citep{witt2009applicability} and
quantum Monte Carlo methods\citep{Mouhat_Casula_zundelQMC_2017,Schmidt_Roy_PIMD-H2O_2018,bertaina_vitali_statistical_2017}
are popular examples. They can be employed for simulations of vibrational
spectra accounting for anharmonicities in both frequency and intensity. 

In this context, semiclassical (SC) Molecular Dynamics (MD) is a powerful
tool for investigating molecular vibrational zero point and excited
eigenenergies. The SC propagator, obtained upon stationary phase approximation
of the parent Feynman's path integral representation, is equivalent
to the short-time propagator proposed by van Vleck.\citep{VanVleck_SCpropagator_1928,Cao_Voth_SCcorrelation_1996}
The original formulation, that mainly suffered from the need to solve
a difficult double boundary problem, was rearranged in a more useful
way first by Miller with his initial value representation (IVR),\citep{Miller_Atom-Diatom_1970,Miller_S-Matrix_1970,Miller_Molecularcollisions_1974,miller2001semiclassical}
and then by the likes of Heller, Herman, Kluk, and Kay that provided
a more manageable representation of the propagator in terms of coherent
states.\citep{Heller_FrozenGaussian_1981,Herman_Kluk_SCnonspreading_1984,Kay_Integralexpression_1994}
However, the original SCIVR requires to deal with a multidimensional
phase-space integration of a real-time oscillatory integrand limiting
the range of applicability of the method. 

\textcolor{black}{Recent advances have permitted to reduce the number
of trajectories required by semiclassical IVR simulations, as well
as the dimensionality of the calculations. Techniques like Filinov
and generalized Filinov filtering,\citep{Makri_Miller_Filinov_1988,Wang_Miller_GeneralizedFilinov_2001,Church_Ananth_Filinov_2017}
and cellular dynamics\citep{Heller_Cellulardynamics_1991,Sulc_Vanicek_CellularDephasing_2012}
were shown effective in speeding up the convergence of the semiclassical
integrations. The same target was achieved by the time averaged version
of SCIVR by Kaledin and Miller,\citep{Kaledin_Miller_Timeaveraging_2003,Kaledin_Miller_TAmolecules_2003}
and by techniques developed in our group like the mixed time-averaging
SCIVR\citep{Buchholz_Ceotto_MixedSC_2016,Buchholz_Ceotto_applicationMixed_2017,Ceotto_Buchholz_SAM_2018}
and the multiple coherent (MC) SCIVR.\citep{Ceotto_AspuruGuzik_Multiplecoherent_2009,Ceotto_AspuruGuzik_PCCPFirstprinciples_2009,Ceotto_Tantardini_Copper100_2010,Ceotto_AspuruGuzik_Curseofdimensionality_2011,Ceotto_AspuruGuzik_Firstprinciples_2011,Conte_Ceotto_NH3_2013,Tamascelli_Ceotto_GPU_2014}
Other methods have been introduced to reduce the dimensionality of
the semiclassical investigation.\citep{Wehrle_Vanicek_Oligothiophenes_2014}
This is also the case of hybrid approaches\citep{Grossmann_SChybrid_2006}
and ``divide-and-conquer'' (DC) SCIVR techniques.\citep{ceotto_conte_DCSCIVR_2017,DiLiberto_Ceotto_Jacobiano_2018} }

\textcolor{black}{The SC hallmark lies on the possibility to perform
quantum dynamics starting from classical trajectories, a feature that
gives SC approaches a clear edge in dealing with hight dimensional
systems. MD is employed to explore the molecular configurational space
including regions away from a specific well, providing a reliable
description of the global surface also for systems characterized by
multiple minima.\citep{Gabas_Ceotto_Glycine_2017,Ceotto_watercluster_18}
Furthermore, differently from other methodologies based on MD, the
classical trajectories are not associated to a target temperature,
so that thermal effects can be added }\textcolor{black}{\emph{a posteriori}}\textcolor{black}{{}
without running a new simulation. However, semiclassical (power) spectra
are calculated from the time evolution of coherent states and peak
amplitudes obtained from these simulations are not necessarily related
to absorption intensities. There are actually some relevant examples
in the literature of semiclassical computations in which absorption
intensities are estimated and compared to the experiments,\citep{Tatchen_Pollak_Onthefly_2009,Wehrle_Vanicek_Oligothiophenes_2014,Batista_Miller_I2anion_1999}
but in approaches using a limited number of trajectories only the
eigenvalues of the spectral decomposition of the vibrational Hamiltonian
are usually obtained from semiclassical dynamics, with the exception
of a recent work in which Ceotto et al. outlined a SC approach able
to get eigenfunctions and applied it to the case of the CO}\textsubscript{\textcolor{black}{2}}\textcolor{black}{{}
molecule.\citep{Ceotto_AspuruGuzik_Firstprinciples_2011} This pre-existing
method has a main drawback though, i.e. it relies on a grid in configurational
space and so it is not suitable for the general treatment of systems
with many degrees of freedom.}

\textcolor{black}{In this work we show how to include the calculation
of vibrational eigenfunctions into the semiclassical formalism by
expanding them on harmonic states. Relying on harmonic states, this
method has the advantage to preserve the description of the properties
of a system in terms of harmonic ingredients allowing to perform,
in an easy and intuitive way, the assignment of each eigenstate (and
absorption peak) in terms of normal modes of vibration. Starting from
the knowledge of semiclassical vibrational eigenfunctions, we will
show how the new formalism can be used to compute, at any temperature,
the IR absorption intensities as well as any other observable that
can be represented as a function of the molecular configurational
space. The method does not require any grid set-ups, it keeps the
possibility of application to high dimensional molecular systems and
completes the general SC treatment of molecular vibrations. }

\textcolor{black}{In the following of this article the methodology
will be derived, tested on a 1D Morse oscillator, and then applied
to determine the eigenfunctions and the IR spectrum of the H}\textsubscript{\textcolor{black}{2}}\textcolor{black}{O
molecule. The paper ends with some conclusions and perspectives.}

\section{Theory\label{sec:Theory}}

\textbf{\textcolor{black}{Notation and preliminary definitions.}}\textcolor{black}{{}
We start by considering the standard representation of a molecular
system within the ground state Born Oppenheimer (BO) adiabatic approximation
in which the wavefunction of nuclei and electrons is decomposed as}

\textcolor{black}{
\begin{align}
\Psi_{n}(\mathbf{r},\mathbf{R})=\varphi_{0}(\mathbf{r};\mathbf{R})e_{n}(\mathbf{R}).\label{eq:Psi-BO}
\end{align}
Eq. (\ref{eq:Psi-BO}) allows to describe the nuclear motion via the
electronic Potential Energy Surface (PES), derived from the ground
state energies of the electronic Hamiltonian at each nuclear structure.
Within this notation, $\textbf{r}$ is the vector collecting position
and spin coordinates of all electrons while $\textbf{R}=\{R_{\alpha}\}_{\alpha=1}^{3N}$
represents a given molecular configuration with index $\alpha\in\{1_{x},~1_{y},1_{z},2_{x},2_{y},\dots,N_{z}\}$.
The adiabatic nuclear motion is governed by the vibrational Hamiltonian
operator }

\textcolor{black}{
\begin{align}
\hat{H}=\hat{T}+\hat{V}\label{eq:BO_Hamiltonian}
\end{align}
with the potential $V(\boldsymbol{R})$ given by the PES. }

\textcolor{black}{If $\textbf{R}_{eq}$ labels a given configuration
of minimum energy on the surface, the second order approximation to
the spectrum of $\hat{H}$ around $\textbf{R}_{eq}$ is derived by
diagonalizing the mass scaled Hessian matrix of the PES in $\textbf{R}_{eq}$}

\textcolor{black}{
\begin{align}
\sum_{\alpha,\beta}\xi_{\beta}^{\gamma}~U_{\beta\alpha}~\xi_{\alpha}^{\lambda}=\omega_{\gamma}^{2}~\delta_{\gamma\lambda},
\end{align}
where $U_{\alpha\beta}=\frac{1}{\sqrt{m_{\alpha}m_{\beta}}}\left.\frac{\partial^{2}V}{\partial R_{\alpha}\partial R_{\beta}}\right|_{\boldsymbol{R}=\boldsymbol{R}_{eq}}$
and $\xi_{\alpha}^{\lambda}$ is the $\alpha^{th}$ component of the
$\lambda^{th}$ eigenvector of $U$. The vibrational Hamiltonian can
be conveniently expressed in terms of mass scaled normal mode coordinates
centered on the equilibrium geometry}

\textcolor{black}{
\begin{align}
\textbf{Q}=\textbf{q}-\textbf{q}_{eq}.
\end{align}
 $\textbf{q}$ is the vector of normal mode coordinates defined as}

\textcolor{black}{
\begin{align}
q_{\alpha}=\sum_{\beta}\xi_{\beta}^{_{\alpha}}~R_{\beta}\sqrt{m_{\beta}}\label{eq:nm}
\end{align}
and $\textbf{q}_{eq}$ is the normal mode vector corresponding to
the equilibrium position $\textbf{R}_{eq}$. The spectral decomposition
of $\hat{H}$ in terms of vibrational bound states}

\textcolor{black}{
\begin{align}
\hat{H}\ket{e_{n}}=E_{n}\ket{e_{n}},\label{eq:H-spectrum}
\end{align}
beyond the harmonic approximation, can be derived using the time propagation
operator
\begin{align}
\mathcal{\hat{{P}}}(t)=e^{-\frac{i}{\hbar}\hat{H}t}=\sum_{n}e^{-\frac{i}{\hbar}E_{n}t}\ket{e_{n}}\bra{e_{n}}\label{eq:propagator-exact}
\end{align}
to compute the recurring time-dependent overlap (also known as the
survival amplitude) of the reference state $|\chi\rangle$}

\textcolor{black}{
\begin{align}
I_{\chi}(t) & \equiv\bra{\chi}\hat{\mathcal{P}}(t)\ket{\chi}=\label{eq:Rec-T-overlap}\\
 & =\sum_{n}e^{-\frac{i}{\hbar}E_{n}t}~\left|\braket{\chi|e_{n}}\right|^{2}\nonumber 
\end{align}
in which the second equality is obtained introducing the representation
of the propagator in the basis of the energy eigenvectors reported
in Eq. (\ref{eq:propagator-exact}). Eq. (\ref{eq:Rec-T-overlap})
implies that, independently of the specific choice for the reference
state $\left|\chi\right\rangle $ to evolve, the eigenenergies of
$\hat{H}$ can be found by taking the peak positions of the power
spectrum of $I_{\chi}(t)$ }

\textcolor{black}{
\begin{flalign}
\tilde{I}_{\textbf{\ensuremath{\chi}}}(E)\!=\frac{1}{2\pi T} & \int_{-T}^{T}\!\!dt\braket{\textbf{\ensuremath{\chi}}|\hat{\mathcal{P}}(t)|\textbf{\ensuremath{\chi}}}e^{\frac{i}{\hbar}Et}=\label{eq:Rec-T-ovelap-power}\\
=\!\frac{1}{\pi T}~Re & \left[\int_{0}^{T}\!\!dt\braket{\chi|\hat{\mathcal{P}}(t)|\chi}e^{\frac{i}{\hbar}Et}\right]\nonumber 
\end{flalign}
}

\noindent \textcolor{black}{where the tilde symbol indicates the action
of the Fourier transform operator. The second equality is simply obtained
considering that $\hat{\mathcal{P}}(t)=\hat{\mathcal{P}}^{\dagger}(-t)$.
Furthermore, and a key point for our purposes, the intensity of each
peak of $\tilde{I}_{\chi}(E_{n})$ is proportional to the square modulus
of the projection of the reference state $|\chi>$ onto the corresponding
eigenstate $|e_{n}>$ as shown in Eq. (\ref{eq:Rec-T-overlap}). These
quantities, in which the quantum propagator is approximated at the
semiclassical level of theory, will be central for the following derivations.}\textbf{\textcolor{black}{}}\\
\textbf{\textcolor{black}{Energy eigenfunctions in a harmonic basis
set.}}\textcolor{black}{{} We can now conveniently consider the complete
and orthonormal N-dimensional basis set $\left\{ \ket{\textbf{\ensuremath{\phi}}_{{\bf K}}}\right\} $
obtained from the Hartree product of one-dimensional harmonic states}

\textcolor{black}{
\begin{align}
\ket{\textbf{\ensuremath{\phi_{\mathbf{K}}}}} & =\ket{\ensuremath{\phi_{\mathbf{K}}^{(1)}},\phi_{\mathbf{K}}^{(2)},~\dots~,\phi_{\mathbf{K}}^{(N_{v})}}=\nonumber \\
 & =\ket{\phi_{\mathbf{K}}^{(1)}}~\dots\ \ket{\phi_{\mathbf{K}}^{(N_{v})}}\label{eq:harm-basis-ket}
\end{align}
where $N_{v}$ is the number of vibrations of the system ($3N-5$
for linear molecules, $3N-6$ otherwise) and}

\textcolor{black}{
\begin{align}
\phi_{\mathbf{K}}^{(\alpha)}(Q_{\alpha}) & =\braket{Q_{\alpha}|\phi_{\mathbf{K}}^{(\alpha)}}=\frac{1}{\sqrt{2^{K_{\alpha}}K_{\alpha}!}}\left(\frac{\omega_{\alpha}}{\pi\hbar}\right)^{\frac{1}{4}}\times\nonumber \\
 & e^{-\frac{\omega_{\alpha}Q_{\alpha}^{2}}{2\hbar}}h_{K_{\alpha}}\left(\sqrt{\frac{\omega_{\alpha}}{\hbar}}~Q_{\alpha}\right).\label{eq:harm-basis-x}
\end{align}
$h_{K_{\alpha}}$ is the $K_{\alpha}^{th}$ order Hermite polynomial.
The vibrational eigenstates of the nuclear Hamiltonian can be expanded
in this basis set, i.e.}

\textcolor{black}{
\begin{align}
\ket{e_{n}}=\sum_{\textbf{K}}C_{n,\mathbf{K}}\ket{\phi_{\mathbf{K}}}\label{eq:en-expansion}
\end{align}
where $C_{n,\mathbf{K}}=\left<\phi_{\mathbf{K}}|e_{n}\right>$ are
real expansion coefficients.}

\textcolor{black}{According to Eq.(\ref{eq:Rec-T-overlap}), the square
modulus of the generic coefficient $C_{n,\textbf{K}}$ can be computed
considering that it is proportional to the intensities of the Fourier
transform of the recurring time-dependent overlap of the corresponding
harmonic state $\phi_{{\bf K}}$ at the eigenvalue of the vibrational
Hamiltonian, i.e.
\begin{align}
\tilde{I}_{\phi_{{\bf K}}}(E_{n})\!=|\braket{\phi_{{\bf K}}|e_{n}}|^{2}= & \frac{1}{\pi T}~Re\left[\int_{0}^{T}\!\!dt\braket{\phi_{{\bf K}}|\hat{\mathcal{P}}(t)|\phi_{{\bf K}}}e^{\frac{i}{\hbar}E_{n}t}\right]\label{eq:coeff_fourier_transf}
\end{align}
and hence}

\textcolor{black}{
\begin{align}
|C_{n,\textbf{K}}|^{2} & =\tilde{I}_{\phi_{{\bf K}}}(E_{n}).\label{eq:coeff-square}
\end{align}
}

\noindent \textcolor{black}{This means that just the sign of the coefficients
$C_{n,\textbf{K}}$ remains undetermined. However, it can be gained
by considering the following time-dependent overlap}

\textcolor{black}{
\begin{align}
I_{\textbf{0}\textbf{K}}(t) & =(\bra{\phi_{\boldsymbol{0}}}+\bra{\phi_{\mathbf{K}}})\hat{\mathcal{P}}(t)(\ket{\textbf{\ensuremath{\phi_{\boldsymbol{0}}}}}+\ket{\textbf{\ensuremath{\phi_{\mathbf{K}}}}})=\nonumber \\
 & =\sum_{n}|\bra{e_{n}}(\ket{\phi_{\boldsymbol{0}}}+\ket{\phi_{\mathbf{K}}})|^{2}e^{-iE_{n}t},\label{eq:I-0J}
\end{align}
}

\noindent \textcolor{black}{where $\ket{\phi_{\boldsymbol{0}}}$ indicates
the harmonic ground state. In fact, by Fourier transforming Eq. (\ref{eq:I-0J})
and using Eq. (\ref{eq:coeff-square}), we get}

\textcolor{black}{
\begin{align}
\tilde{I}_{\textbf{0}\textbf{K}}(E_{n}) & =(C_{n,\boldsymbol{0}})^{2}+(C_{n,\mathbf{K}})^{2}+2~C_{n,\mathbf{K}}C_{n,\boldsymbol{0}}=\nonumber \\
 & =\tilde{I}_{\phi_{{\bf 0}}}(E_{n})+\tilde{I}_{\phi_{{\bf K}}}(E_{n})+2~C_{n,\mathbf{K}}C_{n,\boldsymbol{0}.}\label{eq:coeff-sign}
\end{align}
Solving Eq. (\ref{eq:coeff-sign}) for $C_{n,\mathbf{K}}$ and noting
that $C_{n,\mathbf{0}}=sign(C_{n,\mathbf{0}})\sqrt{\tilde{I}_{\mathbf{\textbf{0}}}(E_{n})}$
leads to the following equation}

\textcolor{black}{
\begin{align}
C_{n,\mathbf{K}}=sign(C_{n,\boldsymbol{0}})\frac{\tilde{I}_{\textbf{0}\textbf{K}}(E_{n})-\tilde{I}_{\phi_{{\bf 0}}}(E_{n})-\tilde{I}_{\phi_{{\bf k}}}(E_{n})}{2\sqrt{\tilde{I}_{\phi_{{\bf 0}}}(E_{n})}},\label{eq:TA-coeff-last}
\end{align}
}

\noindent \textcolor{black}{in which $sign(C_{n,\boldsymbol{0}})=\pm1$
just sets the global sign of $\ket{e_{n}}$ and it is, hence, irrelevant.}

\noindent \textbf{Semiclassical calculation of time recurring overlaps
of harmonic states.} In our SC methodology, we compute the recurring
time-dependent overlap using the following working formula\citep{Ceotto_AspuruGuzik_Multiplecoherent_2009} 

\begin{widetext}

\begin{align}
\tilde{I}_{\chi}^{SC}(E)=\frac{1}{(2\pi\hbar)^{N_{v}}}\frac{1}{2\pi\hbar T}\sum_{j=1}^{N_{s}}\left|\int_{0}^{T}dt\braket{\chi|\mathbf{Q}_{t}^{(j)},\mathbf{p}_{t}^{(j)}}e^{i[S_{t}^{(j)}+Et+\phi_{t}]/\hbar}\right|^{2},\label{eq:MCTA-SC-power}
\end{align}

\end{widetext}where $N_{s}$ is the number of vibrational states
to compute; $\mathbf{Q}_{t}^{(j)}$ and $\mathbf{p}_{t}^{(j)}$ are
the classical normal mode displacement and momentum vectors at time
t obtained propagating the $j^{th}$ classical trajectory with initial
conditions $(\mathbf{Q}_{0}^{(j)},\mathbf{p}_{0}^{(j)})$ under the
effect of the classical vibrational Hamiltonian ; $\ket{\mathbf{Q}_{t}^{(j)},\mathbf{p}_{t}^{(j)}}$
are coherent states of the form

\begin{widetext}

\begin{align}
\braket{\mathbf{x}|\mathbf{Q}_{t},\mathbf{p}_{t}}=\left(\frac{det(\boldsymbol{\gamma})}{\pi}\right)^{\frac{N_{v}}{4}}e^{-\frac{1}{2}(\mathbf{x}-\mathbf{Q}_{t})^{T}\boldsymbol{\gamma}(\mathbf{x}-\mathbf{Q}_{t})+\frac{i}{\hbar}\mathbf{p}_{t}^{T}(\mathbf{x}-\mathbf{Q}_{t})}\label{eq:def-coherents}
\end{align}

\end{widetext}where $\boldsymbol{\gamma}$ is the $N_{v}\times N_{v}$
diagonal matrix, with diagonal elements equal to the harmonic frequencies
$\{\omega_{\lambda}\}_{\lambda=1}^{N_{v}}$; $S_{t}^{(j)}$ is the
classical action at time $t$ computed along the trajectory in the
spirit of Feynman's formulation of path integral quantum mechanics,
and, finally, $\phi_{t}(\mathbf{Q}_{0},\mathbf{p}_{0})$ is the phase
of $C_{t}(\mathbf{Q}_{0},\mathbf{p}_{0})$, the Herman-Kluk prefactor
at time $t$ that accounts for second order quantum fluctuations around
each classical path and which is obtained as\citep{Herman_Kluk_SCnonspreading_1984,DiLiberto_Ceotto_Prefactors_2016}

\begin{widetext}

\begin{equation}
C_{t}\left(\mathbf{Q}_{0},\mathbf{p}_{0}\right)=\sqrt{\frac{1}{2^{N_{v}}}\left|\frac{\partial\mathbf{Q}_{t}}{\partial\mathbf{Q}_{0}}+\frac{\partial\mathbf{p}_{t}}{\partial\mathbf{p}_{0}}-i\hbar\boldsymbol{\gamma}\frac{\partial\mathbf{Q}_{t}}{\partial\mathbf{p}_{0}}+\frac{i\boldsymbol{\gamma}^{-1}}{\hbar}\frac{\partial\mathbf{p}_{t}}{\partial\mathbf{Q}_{0}}\right|}.\label{eq:prefactor}
\end{equation}

\end{widetext}

The starting point of our SC implementation is the Herman-Kluk propagator
in its time averaged version by Kaledin and Miller,\citep{Kaledin_Miller_Timeaveraging_2003}
that can be used to compute the Fourier transformed time-dependent
recurring overlaps ($\tilde{I}_{{\bf \chi}}$) but which requires
to perform a multidimensional integration over initial conditions
in phase space. This is usually achieved by means of Monte Carlo techniques
and the method has been applied successfully to describe the vibrational
properties of several molecules, yielding very accurate results upon
evolution of about $10^{3}$ trajectories per degree of freedom.\citep{Kaledin_Miller_TAmolecules_2003}
In our approach, the computational overhead required to construct
the quantum propagator is enormously decreased. In fact, as indicated
in Eq. (\ref{eq:MCTA-SC-power}), we follow the footsteps of the multiple
coherent technique\citep{Ceotto_AspuruGuzik_Multiplecoherent_2009}
and, rather than relying on a full Monte Carlo sampling of the phase
space, the propagator is constructed using only $N_{s}$ tailored
trajectories, i.e. one for each target vibrational state. These trajectories
are selected and derived carefully on the basis of the starting harmonic
approximation to the Hamiltonian. In particular, the initial position
is selected to be in the minimum of the potential (equilibrium position),
while the initial velocities are chosen in a way to assign to each
normal mode a content of kinetic energy equal to $T_{\alpha}=(n_{\alpha}+\frac{1}{2})\hbar\omega_{\alpha}$.
Each trajectory requires to be evolved for a very short time (1-2
ps) without any preliminary equilibration to be performed. 

Apart from the evolution of the classical trajectories, calculation
of $\tilde{I}_{\chi}^{SC}(E)$ using Eq. (\ref{eq:MCTA-SC-power})
requires to evaluate also the phase of the prefactor reported in Eq.
(\ref{eq:prefactor}). The prefactor depends on the stability matrix
elements $\frac{\partial(Q_{t}^{(\alpha)},p_{t}^{(\alpha)})}{\partial(Q_{0}^{(\alpha)},p_{0}^{(\alpha)})}$,
which are obtained via numerical integration of their symplectic equations
of motion along the classical trajectory.\citep{Brewer_Manolopoulos_15dof_1997}
For this purpose, however, the computationally-expensive calculation
of the instantaneous Hessian matrix $\left.\frac{\partial^{2}V}{\partial Q_{\alpha}\partial Q_{\beta}}\right|_{\mathbf{Q}_{t}}$
is needed. Specific algorithms have been developed to ease computational
costs in high dimensional applications by reducing the number of Hessian
calls.\citep{Zhuang_Ceotto_Hessianapprox_2012,Ceotto_Hase_AcceleratedSC_2013}
The reference state to evolve is usually chosen to be in the form
of a coherent state $\ket{\chi}=\ket{\bar{\mathbf{Q}},\bar{\mathbf{p}}}$
so that the scalar product $\braket{\chi|\mathbf{Q}_{t},\mathbf{p}_{t}}$
can be computed analytically at each phase space point visited during
the classical MD. However, for our purposes, we want to consider the
case in which $\ket{\chi}=\ket{\phi_{\mathbf{K}}}$. The calculation
of the following overlap is then needed: 

\begin{align}
\braket{\mathbf{\phi_{\mathbf{K}}}|\mathbf{Q}_{t},\mathbf{p}_{t}}=\prod_{\alpha=1}^{N_{v}}\braket{\phi_{\mathbf{K}}^{(\alpha)}|Q_{t}^{(\alpha)},p_{t}^{(\alpha)}},\label{eq:ska-harm-cohe-vec}
\end{align}
 which has the analytical form

\begin{align}
\braket{\phi_{\mathbf{K}}^{(\alpha)}|Q_{t}^{(\alpha)},p_{t}^{(\alpha)}}=e^{-\frac{i}{2\hbar}Q_{t}^{(\alpha)}p_{t}^{(\alpha)}} & \times\nonumber \\
\times e^{-\frac{\omega_{\alpha}}{4\hbar}[Q_{t}^{(\alpha)}]^{2}-\frac{1}{4\omega_{\alpha}\hbar}[p_{t}^{(\alpha)}]^{2}} & \times\nonumber \\
\times\frac{(\sqrt{\frac{\omega_{\alpha}}{2\hbar}}~Q_{t}^{(\alpha)}~+~i~\sqrt{\frac{1}{2\omega_{\alpha}\hbar}}~p_{t}^{(\alpha)})^{K_{\alpha}}}{\sqrt{K_{\alpha}!}}\label{eq:harm-coherent-overlap}
\end{align}

The details of the analytical derivation of Eq. \ref{eq:harm-coherent-overlap}
can be found in Appendix A.

\textbf{Calculation of temperature dependent Absorption Spectra.}
Once the spectral decomposition of the vibrational operator has been
achieved, the IR absorption intensities can be obtained using quantum
linear response theory in its sum-over-state version. Within this
formalism, the IR spectrum for isotropic and homogeneous molecular
systems is written as\citep{McQuarrie_absorptionspectrum_1976}

\begin{align}
S(\omega,T)=\sum_{n\neq m}[P_{n}(T)-P_{m}(T)]F_{nm}\delta(\omega-\Omega_{nm})\label{eq:abs-spec}
\end{align}
where $\Omega_{nm}=E_{m}-E_{n}$ is the difference between vibrational
excitation energies, $P_{n}=e^{-\frac{E_{n}}{k_{B}T}}/Z$ is the $n^{th}$
vibrational state population at a given temperature T (with $Z=\sum_{n}e^{-\frac{E_{n}}{k_{B}T}}$
being the partition function), and

\begin{align}
 & F_{nm}\propto\Omega_{nm}|\braket{\Psi_{n}|\hat{\mu}|\Psi_{m}}|^{2}=\Omega_{nm}\times\label{eq:osc-str-R-r}\\
\times & \!\!\int\!\!d\mathbf{R}\!\!\int\!\!d\mathbf{r}|\varphi_{0}(\mathbf{r};\mathbf{R})|^{2}e_{n}(\mathbf{R})e_{m}(\mathbf{R})\mu(\mathbf{r},\mathbf{R})\nonumber 
\end{align}

\noindent are the oscillator strengths, which depend on the full Hamiltonian
eigenstates. In the second equality of Eq.(\ref{eq:osc-str-R-r})
we used the BO approximation for the total wavefunction as illustrated
in Eq. (\ref{eq:Psi-BO}) with the dipole function $\mu$ that can
be decomposed as

\begin{align}
\mu(\mathbf{r},\mathbf{R})~ & =\sum_{\alpha}Z_{\alpha}R_{\alpha}+e\sum_{i}r_{i}=\nonumber \\
 & =~~\mu_{N}(\mathbf{R})~+~\mu_{e}(\mathbf{r})\label{eq:mu-R-r}
\end{align}
where $Z_{\alpha}\in\{Z_{1},Z_{1},Z_{1},\dots,Z_{N},Z_{N},Z_{N}\}$
is the charge associated to the $\alpha^{th}$ degree of freedom of
the system and $e$ is the charge of the electron. The separable form
of the dipole operator in Eq. (\ref{eq:mu-R-r}) allows to compute
the oscillator strengths as matrix elements over vibrational states,
i.e.

\begin{align}
F_{nm}~\propto & ~\Omega_{nm}~|<~e_{n}~|~\hat{\mu}_{0N}~|~e_{m}>|^{2}\label{eq:osc-str-e}
\end{align}
where

\begin{align}
\hat{\mu}_{0N}(\mathbf{R}) & =\hat{\mu}_{N}(\mathbf{R})+\hat{\mu}_{e0}(\mathbf{R})\label{eq:op-dip-tot-R}
\end{align}
and

\begin{align}
\mu_{e0}(\mathbf{R})~=~\int d\mathbf{r}~|\varphi_{0}(\mathbf{r};\mathbf{R})|^{2}\mu_{e}(\mathbf{r})\label{eq:op-dip-e0-R}
\end{align}
is the electronic dipole associated to a given nuclear configuration.
Using Eq. (\ref{eq:osc-str-e}), the absorption intensities can be
obtained by computing the following integral over the nuclear configurational
space:

\begin{align}
M_{nm}=\int d\mathbf{Q}~e_{n}(\mathbf{Q})e_{m}(\mathbf{Q})\mu_{0N}(\mathbf{Q})\label{eq:trans-dip-integral}
\end{align}
 the only unknown term being the electronic dipole of Eq. (\ref{eq:op-dip-e0-R})
that demands for an electronic structure calculation at every nuclear
configuration. 

Calculation of these integrals can be approached through a Monte Carlo
sampling. This can be done by taking advantage from the fact that,
in the expansion of vibrational eigenstates in the harmonic basis
of Eq. (\ref{eq:harm-basis-x}), the Gaussian term (present in each
harmonic function) can be factorized out leading to

\begin{align}
e_{n}(\mathbf{Q})=G(\mathbf{Q},\pmb{\omega})\sum_{{\bf K}}C_{n,{\bf K}}~\bar{{\bf \phi}}_{\mathbf{K}}(\mathbf{Q})
\end{align}
where $G(\mathbf{Q},\pmb{\omega})=e^{-\frac{1}{2\hbar}~\mathbf{Q}^{T}\pmb{\omega}~\mathbf{Q}}$
is the $N_{v}$-dimensional Gaussian term and \\
$\bar{{\bf \phi}}_{\mathbf{K}}(\mathbf{Q})=\prod_{\alpha=1}^{N_{v}}\frac{1}{\sqrt{2^{K_{\alpha}}K_{\alpha}!}}~\left(\frac{\omega_{\alpha}}{\pi\hbar}\right)^{\frac{1}{4}}h_{K_{\alpha}}\left(\sqrt{\frac{\omega_{\alpha}}{\hbar}}~Q_{\alpha}\right)$
is the coordinate representation of the harmonic state $|\phi_{\mathbf{K}}>$
without the Gaussian terms. The integral in Eq. (\ref{eq:trans-dip-integral})
can be conveniently recast in the following way:

\begin{align}
M_{nm}=\int\left[d\mathbf{Q}~G(\mathbf{Q},2\pmb{\omega})\right]\bar{e}_{nm}(\mathbf{Q})~\hat{\mu}_{0N}(\mathbf{Q})\label{eq:trans-dip-integral-for-MC}
\end{align}
where

\begin{align}
\bar{e}_{nm}(\mathbf{Q})=\left(\sum_{\mathbf{K}}C_{n,\mathbf{K}}\bar{{\bf \phi}}_{\mathbf{K}}(\mathbf{Q})\right)\left(\sum_{\mathbf{K'}}C_{m,\mathbf{K'}}\bar{{\bf \phi}}_{\mathbf{K'}}(\mathbf{Q})\right)
\end{align}
Written as in Eq. (\ref{eq:trans-dip-integral-for-MC}), this integral
is particularly well suited for Monte Carlo sampling. In fact Gaussian
distributions can be easily generated by means of the Box-Muller algorithm,\citep{Numerical_Recipes}
so that the integrals can be evaluated as

\begin{align}
M_{nm}=\mathcal{K}\lim_{N_{MC}\rightarrow\infty}\frac{1}{N_{MC}}\sum_{k=1}^{N_{MC}}\bar{e}_{nm}(\mathbf{Q}_{k})~\hat{\mu}_{0N}(\mathbf{Q}_{k})\label{eq:trans-dip-integral-with-BM-MC}
\end{align}
where $\left\{ \mathbf{Q}_{k}\right\} $ is a set of molecular configurations
generated along the multivariate Gaussian distribution $\mathcal{N}(\mathbf{Q},\sqrt{\frac{1}{2}\pmb{\omega}^{-1}})$
and

\begin{align}
\mathcal{K}=\frac{G(\mathbf{Q},2\pmb{\omega})}{\mathcal{N}(\mathbf{Q},\sqrt{\frac{1}{2}\pmb{\omega}^{-1}})}=\sqrt{\frac{(2\pi)^{N_{v}}}{2\left|\pmb{\omega}\right|}}
\end{align}

\section{results and discussions}

\textbf{1D Morse oscillator: }A good test of performances, accuracy
and features of our methodology is represented by the 1D Morse oscillator.
In fact, this model system, even if very simple, contains the level
of anharmonicity that is typically encountered in the description
of molecular bond stretchings. Within the Morse potential functional
form $V(Q)=D_{e}\left[1-e^{-\sqrt{\omega^{2}/2D_{e}}Q}\right]^{2}$,
we set $\omega=0.020~a.u.$ and $D_{e}=0.174~a.u.$ with the aim to
mimic the bond vibration of the H\textsubscript{2} molecule. Five
classical trajectories $\{Q_{t}^{(n)},p_{t}^{(n)}\}_{n=0,..,4}$ were
propagated with initial conditions $Q_{0}^{(n)}=0$ and $p_{0}^{(n)}=\sqrt{(2n+1)\omega}$
in order to describe the first five vibrational states (ground state
plus first four excited states) with our SC propagator. These trajectories
have been obtained via numerical integration of the classical equations
of motion directly in normal modes and using a fourth order symplectic
numerical integrator.\citep{Brewer_Manolopoulos_15dof_1997} Gradients
and Hessians were computed numerically through central finite difference
formulae.\citep{forn98} 

The generic $n^{th}$ vibrational eigenenergy has been obtained following
the prescription of Eq.(\ref{eq:MCTA-SC-power}), calculating the
Fourier transform of the time recurrent overlap of the $n^{th}$ harmonic
state and extracting the frequency that corresponds to the $n^{th}$
peak position (results are shown in Table S1 in the supplemental material).
The semiclassical values thus obtained are very close to the reference
analytical values $E_{n}=(n+\frac{1}{2})\hbar\omega-[\hbar\omega(n+\frac{1}{2})]^{2}/4D_{e}$,
with errors of the order of the wavenumber up to the second excited
state, and of a few tens of $cm^{-1}$ for higher energy states. The
Morse eigenfunctions were expanded in terms of the first 10 harmonic
eigenstates using Eq. \ref{eq:TA-coeff-last} and compared with their
analytical expression: $e_{n}(Q)=N_{n}~z(Q)^{\lambda-n-\frac{1}{2}}~e^{-\frac{1}{2}z}~L_{n}^{\alpha}(z(Q))$,
where $\lambda=\frac{2D_{e}}{\hbar\omega}$, $z(Q)=2\lambda~e^{-y(Q)}$,
$y(Q)=\sqrt{\frac{\omega^{2}}{2D_{e}}}~Q$, $\alpha=2\lambda-2n-1$,
$L_{n}^{\alpha}(z)$ is the generalized Laguerre polynomial, and $N_{n}$
their normalization constants (derived by means of numerical integration). 

\begin{figure}[H]
\centering{}\includegraphics[scale=0.55]{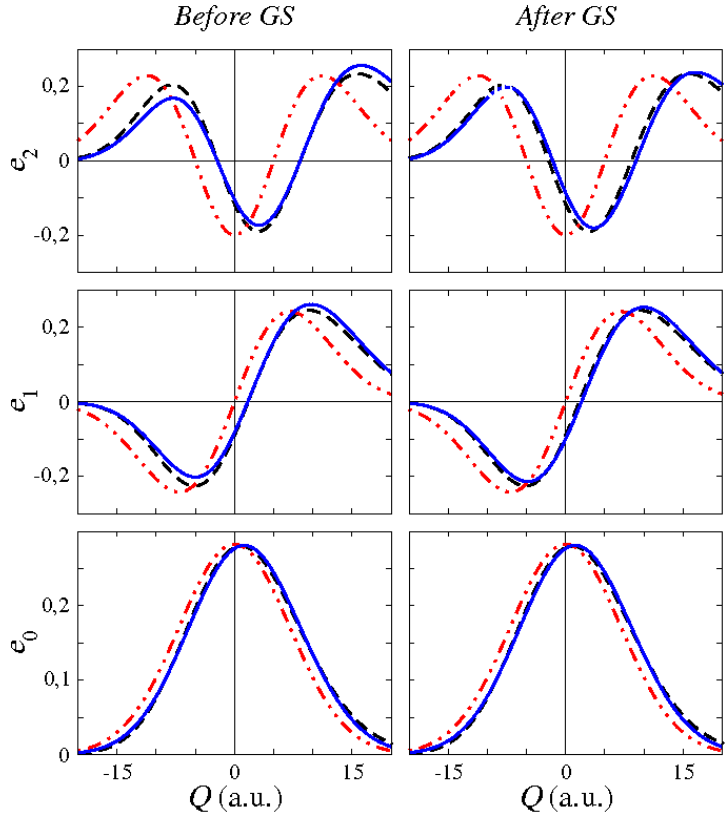} \caption{Semiclassical eigenfunctions (blue continuous lines) for the ground,
first, and second excited state of a 1D-Morse potential upon expansion
on a basis set made of the first 10 harmonic eigenstates. Results
are compared with the corresponding harmonic (red double-dot dashed
lines) and exact (black dashed lines) ones. On the right side column
the wave functions are refined using the Gram Schmidt algorithm while
on the left hand side they are reported before the \textquotedbl a
posteriori\textquotedbl{} orthogonalization. The mass-scaled coordinates
are in a.u.}
\label{fig:wfns-Morse1D-TA} 
\end{figure}

On the left column of Fig. \ref{fig:wfns-Morse1D-TA} we report the
first three SC vibrational eigenfunctions together with the corresponding
exact and harmonic wavefunctions. As for the ground state eigenfunction,
it is evident that the anharmonic corrections are minor and the harmonic
approximation already provides a realistic guess. However, it fails
in locating the maximum of the wavefunction, which shifts from the
harmonic estimate ( $Q=0$) to $Q\sim3$ a.u. in mass-scaled coordinates
or about 0.05 $\lyxmathsym{\AA}$ in cartesian coordinates in the
direction of bond cleavage when the anharmonicity of the potential
is properly accounted for. Interestingly, this effect is already correctly
described when truncating the harmonic basis set at the level of the
first excited state. The corresponding coefficient ($C_{0,1}$ in
our notation) has an amplitude of $\sim0.1$ and the sign of this
coefficient gives the direction of the shift. If a bigger harmonic
basis set is employed, only two other coefficients provide a non-negligible
but minor contribution of the order of the percent. 

The effects of anharmonicity become more and more relevant as the
vibrational energy increases. SC dynamics performs better than the
harmonic approximation in all cases but, starting from $e_{2}(Q)$,
discrepancies between SC and exact wavefunctions become evident (see
Fig. S1 in the supplemental material). This drawback can be eased
by imposing the orthonormalization condition to the SC wavefunctions
after their basic, preliminary estimate obtained via Fourier transform
of the recurring overlap. This can be done efficiently using the Gram
Schmidt (GS) algorithm. Application of the GS scheme is straightforward
and constitutes a simple post-processing refinement of the results.
Furthermore, by construction, the GS algorithm does not manipulate
the ground state wavefunction where the anharmonic corrections are
minor and efficiently accounted for. Wavefunctions are then improved
iteratively starting from the ground state in a way that orthogonality
is enforced only against wavefunctions already optimized. As demonstrated
by the right column of Fig. \ref{fig:wfns-Morse1D-TA}, the procedure
permits to escalate the quality of the SC wavefunctions which is now
not only excellent up to $e_{2}(Q)$, but also very good for $e_{3}(Q)$
and $e_{4}(Q)$ (as reported in Fig. S1 in the supplemental material).

\begin{widetext}

\begin{table}[H]
\begin{centering}
\caption{1D-Morse oscillator dipoles (in a.u.) for selected (non vanishing)
vibrational transitions. The transitions are reported in the first
column. Numerical estimates obtained using our method are reported
before and after the application of the GS procedure and are compared
with their analytical values derived from Eq. \ref{eq:Morse-dipoles-1}
. \label{tab:Morse-dipoles}}
\par\end{centering}
\centering{}%
\begin{tabular}{ccccccc}
 & \phantom{aaa} &  & \phantom{a} &  & \phantom{a} & \tabularnewline
Transition Dipoles &  & Before GS  &  & After GS  &  & \tabularnewline
 &  & SC &  & SC &  & exact\tabularnewline
 &  &  &  &  &  & \tabularnewline
\hline 
\hline 
 &  &  &  &  &  & \tabularnewline
$d_{01}$  &  & 0.16  &  & 0.16  &  & 0.17 \tabularnewline
$d_{12}$  &  & 0.25  &  & 0.24  &  & 0.25 \tabularnewline
$d_{13}$ &  & 0.01  &  & 0.05  &  & 0.04 \tabularnewline
$d_{23}$ &  & 0.34  &  & 0.30  &  & 0.31 \tabularnewline
$d_{24}$ &  & 0.01  &  & 0.07  &  & 0.06 \tabularnewline
$d_{34}$ &  & 0.42  &  & 0.35  &  & 0.36 \tabularnewline
 &  &  &  &  &  & \tabularnewline
\hline 
\end{tabular} 
\end{table}

\end{widetext}

A quantitative estimate of the accuracy reached with our method has
been obtained with the calculation of selected non vanishing nuclear
transition dipoles $d_{nm}=\braket{e_{n}|Q|e_{m}}$. We computed them
numerically on a uniform grid of $10^{4}$ points using the SC eigenfunctions
and compared the results with exact analytical values given by the
following Eq. (\ref{eq:Morse-dipoles-1})\begin{widetext}

\begin{equation}
d_{mn}^{(ex)}=\frac{2(-1)^{m-n+1}}{(m-n)(2K-n-m)}\sqrt{\frac{(K-n)(K-m)~\Gamma(2K-m+1)m!}{\Gamma(2K-n+1)n!}},\label{eq:Morse-dipoles-1}
\end{equation}

\end{widetext} where $K=\lambda-\frac{1}{2}$. Results are reported
in Table \ref{tab:Morse-dipoles}. Transition dipoles $0\rightarrow1$
and $1\rightarrow2$ are accurate within a tolerance of 0.01 a.u.
even before the GS refinement, confirming that the quality of the
ground state and first two excited SC wavefunctions is very high.
The dipoles associated to excitations involving higher energy states
(1 $\rightarrow$ 3, 2 $\rightarrow$ 3, 2 $\rightarrow$ 4, 3 $\rightarrow$
4) are instead less accurate, and SC results are correct only within
a tolerance of 0.05 a.u. However, after GS orthonormalization, the
quality of these exotic transition dipoles improves and their accuracy
becomes comparable to that of dipoles involving lower energy states.

\textbf{H}\textsubscript{\textbf{2}}\textbf{O Molecule: }We now
move to apply our method to the description of the vibrations of the
non-rotating water molecule in vacuum. We employed in our calculation
a pre-existing analytical PES based on a quartic force field involving
the displacement coordinates of the internal angle and the two bonds.\citep{Dressler_Thiel_WaterPES_1997}
First, the Hessian matrix has been diagonalized to find the three
harmonic frequencies of vibration, which are well known to be related
to the symmetric stretch ($\omega_{s}=3831$ cm\textsuperscript{-1}),
the bending ($\omega_{b}=1650$ cm\textsuperscript{-1}), and the
asymmetric stretch ($\omega_{a}=3941$ cm\textsuperscript{-1}) motions.
Consistently to the case of the Morse oscillator, five classical trajectories,
corresponding to the harmonic states (in increasing order of energy)
(0,0,0), (0,1,0), (0,2,0), (1,0,0), (0,0,1), have been run to determine
the 5 lowest-lying vibrational states of water semiclassically. These
trajectories have been evolved by means of the same symplectic numerical
integrator adopted for the Morse oscillator with gradients and Hessians
calculated through the usual central finite-difference scheme. 

Reference quantum molecular dynamics calculations were carried out
by means of the Grid Time-Dependent Schroedinger Equation (GTDSE)
computational package.\citep{ssf+09a} The GTDSE code includes an
implementation of the Lanczos algorithm \citep{lanczos,wyatt95,gc02}
that we exploited to extract the eigenfunctions and eigenvalues of
the Hamiltonian. The space of coordinates was discretized and finite
difference methods \citep{forn98} were employed to calculate the
derivatives required by the Laplacian operator. The accuracy of the
calculations depends on the density of the grid of points in the discretization
and on the number of points (stencil) used to calculate the derivatives.
The finite difference scheme is formally equivalent to discrete variable
representation (Sinc-DVR) methods when including all the grid points
in the stencil,\citep{guantesfarantos99} although this is usually
unnecessary since the convergence is rapidly reached. This feature
makes the GTDSE particularly efficient in the case of a (generalized)
Cartesian coordinate system, where the second derivatives in the Laplacian
return a sparse Hamiltonian matrix. In order to better compare these
DVR results with SC ones for H$_{2}$O, we used the Lanczos algorithm
and represented both the PES and the wavefunctions directly in the
same normal mode coordinates as the semi-classical calculations, with
grid limits $L_{i}=\pm75.0\text{ a.u.}~(i=1,2,3)$ for the mass-scaled
$\boldsymbol{Q}_{i}$ coordinates, using 150 grid points and 15 points
in the stencil along each direction. An additional benchmark calculation,
this time in Jacobi coordinates, was used to extract the DVR vibrational
reference values for the non-rotating molecule (${\bf J}=0$). These
eigenenergies, together with more details about this system of coordinates,
are reported in the supplemental material (see table S12). 

\begin{figure}[H]
\centering{}\includegraphics[bb=37bp 19bp 707bp 546bp,scale=0.5]{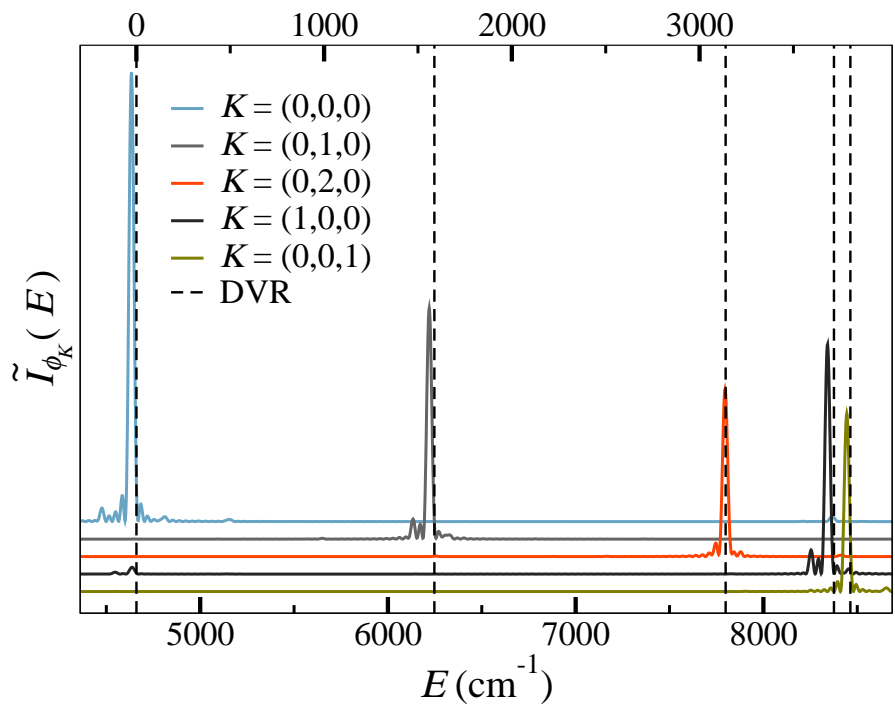}
\caption{Fourier transforms of the recurring time dependent overlaps ($\tilde{I}_{\mathbf{\phi_{{\bf K}}}}(E)$)
obtained evolving the five least energetic harmonic states with the
SC propagator, constructed using a single trajectory with the corresponding
harmonic energy. The power spectra are shifted in the ordinate axis
in order to facilitate visualization and reported in different colors
(indicated in the legend). Reference DVR energy values for the normal
coordinates system are presented with dashed vertical black lines.
The upper horizontal axis reports the shift in frequency (cm\protect\textsuperscript{-1})
from the ZPE peak. }
\label{fig:Itilde-PES-H2O} 
\end{figure}

In Fig. \ref{fig:Itilde-PES-H2O} we report the plots of $\tilde{I}_{\mathbf{\phi_{{\bf K}}}}(E)$
for the first five harmonic states of water. The semiclassical vibrational
energies derived from the positions of the peaks lay within $\sim30~cm^{-1}$
of the reference DVR estimates on the same PES and normal coordinates
system. A similar level of accuracy was obtained \textcolor{black}{by
Kaledin and Miller }propagating coherent states.\citep{Kaledin_Miller_Timeaveraging_2003}
This result confirms the quality of the MC-SCIVR approximation independently
of the particular reference state (harmonic or coherent) propagated.

For a general $N_{v}$-dimensional system, the size of the (truncated)
harmonic basis set, obtained considering all the possible $N_{v}$-dimensional
direct products of 1-dimensional harmonic eigenstates up to the quantum
number $k_{max}$, is $(k_{max}+1)^{N_{v}}$ , and hence it grows
exponentially with the number of vibrational degrees of freedom. This
issue makes the description of vibrational wavefunctions of medium-size
or larger molecular systems (i.e. when $N_{v}\sim10$ or bigger) virtually
undoable because the dimension of the basis set would be too large
to be stored in a computer. This is not the case for the water molecule
($N_{v}=3$) and hence, expanding the eigenfunctions in terms of the
first 11 harmonic states $(k_{max}=10)$ the total number of states
in the basis set adds up to just $11^{3}=1331$. Even if such a calculation
for water is feasible, in view of future applications of this method
to molecules of higher dimensionality we reduced the amount of data
to be stored by setting all coefficients with amplitude smaller than
$0.01$ to zero. The surviving coefficients were refined by enforcing
orthonormality by means of the GS algorithm. In this way, the ground
state eigenfunction was decomposed on just five harmonic states, while
excited states required to increase the basis set size up to about
$10$ elements. 

\begin{figure}[H]
\begin{centering}
\includegraphics[angle=270,scale=0.6]{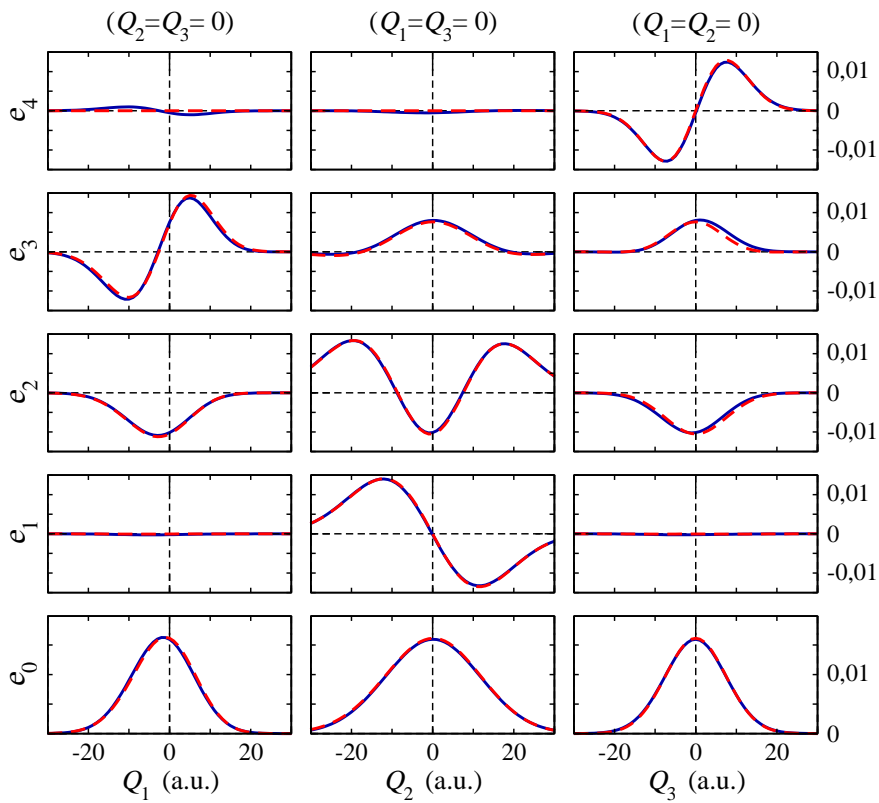}
\par\end{centering}
\caption{Selected cuts of the water eigenfunctions obtained by means of DVR
(in red) and with SC dynamics (in blue). The cuts are performed on
the three directions of the configurational space derived by fixing
two of the three normal coordinates at their equilibrium values. Wavefunctions
are presented in ascending order of energy (from bottom to top) for
states (0,0,0), (0,1,0), (0,2,0), (1,0,0), and (0,0,1).\label{fig:wfns-H2O-SC-DVR-1}}
\end{figure}

The eigenfunctions are plotted in three different cuts of the configurational
space in Fig. \ref{fig:wfns-H2O-SC-DVR-1} where they are also compared
with their reference DVR estimate. The accuracy obtained is very high
for all cases and the effect of the truncation of the basis set is
barely visible on the nodal planes where the SC wavefunctions are
slightly overstructured. In perspective, this procedure can help overcome
the curse of dimensionality given by the exponential growth of the
size of the harmonic basis as a function of the system dimensionality.
In fact, in order to moderate the number of harmonic states to generate,
a polynomial growth can be enforced for instance by building an initial
basis set which includes only states with a maximum of simultaneously
excited degrees of freedom smaller than $N_{v}$. \textcolor{black}{An
alternative approach would consist in selecting the harmonic states
in the basis set under a constraint on the total energy, which has
to be close to the desired target energy. Then, the same procedure
adopted for H}\textsubscript{\textcolor{black}{2}}\textcolor{black}{O
can be applied on this basis set of reduced size. }

\begin{figure}[H]
\centering{}\includegraphics[scale=0.55]{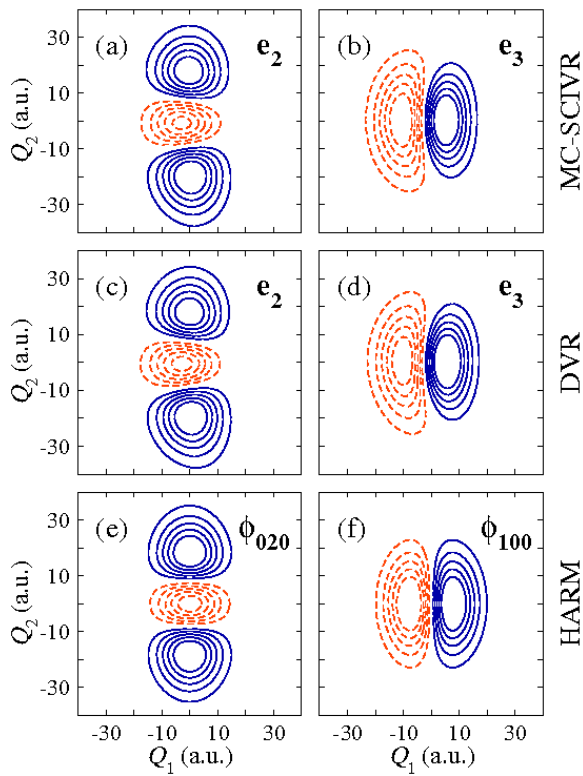} \caption{Bidimensional contour plots of H\protect\textsubscript{2}O vibrational
eigenfunctions obtained setting $Q_{3}=0$. The pristine harmonic
states $\phi_{020}$ and $\phi_{100}$ are reported on the bottom
row. They are the harmonic approximation to the quantum states $e_{2}$
and $e_{3}$ plotted in the first (MC-SCIVR estimate) and second row
(DVR reference).\label{fig:wfns-H2O-SC-DVR}}
 
\end{figure}

Anharmonicity effects for the bending and asymmetric stretching modes
are small because all odd order terms in the PES force field vanish
for symmetry. However, two main effects that characterize the H\textsubscript{2}O
molecule are efficiently accounted for by this method. The first one
involves the symmetric stretch mode along which the potential is approximately
the sum of two Morse-like 1D potentials for the OH bond stretching.
Consistently with the case of the 1D Morse oscillator, the anharmonicity
generates a coefficient $C_{0,(100)}$ of the order of $10~\%$ in
the expansion of the ground state wavefunction (see Table S7 in the
Supplemental Material), shifting the maximum with respect to the harmonic
eigenfunction in the direction of the dissociation. This effect is
clearly visible in Fig. \ref{fig:wfns-H2O-SC-DVR}. 

The second relevant anharmonicity effect regards the presence of a
Fermi resonance between the harmonic states corresponding to a double
excitation of the bending (i.e. state $\phi_{020}$) and the first
excited state for the symmetric stretch $\phi_{100}$. In terms of
expansion coefficients this effect is pointed out by the presence
of two non-negligible terms $C_{2,(100)}\sim C_{3,(020)}\sim0.15$
for the second and third anharmonic states (see table S9 and S10 in
the Supplemental Material). The Fermi resonance is well represented
in Fig. \ref{fig:wfns-H2O-SC-DVR}. Its effect on the shape of the
anharmonic eigenfunction $e_{2}$ is clearly visible in the bidimensional
contour plot reported in panel (a). In fact there is a distortion
of the harmonic symmetry in the direction of negative $Q_{2}$ values
where, as shown in (f), the harmonic $\phi_{100}$ wavefunction is
positive. The anharmonic distortions of state $e_{3}$ due to the
Morse-like shape of the bond stretch potential are also visible in
panel (b), which has a broader decay in the direction of dissociation
(negative $Q_{2}$), opposite to the steeper decay in the direction
of positive $Q_{2}$, where the potential grows more rapidly because
of the repulsive interactions. Excellent overall agreement of the
SC wavefunctions (panels (a) and (b)) with their DVR counterparts
(panels (c) and (d)) is confirmed also by this plot. 

This qualitative agreement between SC and DVR eigenfunctions has been
confirmed quantitatively by computing the oscillator strengths by
means of Eq. \ref{eq:osc-str-e}. The total dipole moment and the
wavefunctions were integrated along the grid of normal mode coordinates.
The nuclear component of the dipole moment is readily available, while
the electronic part was extracted from the fitted dipole surface of
Lodi \emph{et al.}\citep{tennyson_diph2o} The surface was built in
a way that the two components of the electronic dipole moment are
returned parallel and perpendicular to the bond-angle bisector vector,
while the oxygen atom is set at the origin of the reference frame.
For a correct evaluation of the total dipole moment, the electronic
and nuclear dipole contributions had to be calculated in the same
frame and using the same pole. Results, reported in Tab. \ref{tab:h20-os-str-grid-MC},
confirm the high level of accuracy obtained with the SC wavefunctions
with a tolerance of the same order of the one obtained for 1D Morse
potential. Interestingly, due to the presence of the Fermi resonance,
the DVR oscillator strength of the bending overtone transition (F\textsubscript{02})
is of the order of $1\%$ of the fundamental symmetric stretch transition
($F_{03}$) and not exactly zero (as it is in the harmonic approximation)
since there is a minimal contribution coming from the harmonic dipole
$\braket{\phi_{000}|\hat{\mu}|\phi_{100}}$. This effect, however,
is too small to be observed within our SC method because the amplitude
of $F_{02}$ is smaller than the tolerance in the SC estimates. 

\begin{widetext}

\begin{table}[H]
\begin{centering}
\caption{The numerical values obtained for the oscillator strengths of water
evaluated on the DVR grid are reported in the second and third column
using respectively SC and DVR eigenfunctions, respectively. Monte
Carlo estimates, obtained by employing the SC eigenfunctions to generate
molecular configurations in the MC scheme depicted in Eq. \ref{eq:trans-dip-integral-for-MC},
follow in columns 4 and 5. The MC values are reported after the evaluation
of the molecular dipole on 25000 and 50000 structures. For these cases
the statistical error (estimated as the square root of the variance)
is reported in parentheses.\label{tab:h20-os-str-grid-MC}}
\par\end{centering}
\centering{}%
\begin{tabular}{ccccccccc}
Oscillator Strength & \phantom{aa} & $SC$ & \phantom{a} & $DVR$ & \phantom{a} & $MC$ & \phantom{a} & $MC$\tabularnewline
 &  & Grid &  & Grid &  & $25000$ steps &  & $50000$ steps\tabularnewline
 &  &  &  &  &  &  &  & \tabularnewline
\hline 
\hline 
 &  &  &  &  &  &  &  & \tabularnewline
$F_{01}$ &  & 19.3  &  & 19.6 &  & 17.9 ($\pm$2.6)  &  & 18.5 ($\pm$1.9) \tabularnewline
 &  &  &  &  &  &  &  & \tabularnewline
$F_{02}$ &  & 0.01  &  & 0.08 &  & \phantom{1}0.1 ($\pm$0.3)  &  & \phantom{1}0.1 ($\pm$0.2) \tabularnewline
 &  &  &  &  &  &  &  & \tabularnewline
$F_{03}$ &  & 7.0  &  & 7.1 &  & \phantom{1}6.4 ($\pm$2.8)  &  & 5.84 ($\pm$1.9) \tabularnewline
 &  &  &  &  &  &  &  & \tabularnewline
$F_{04}$ &  & 8.81 &  & 8.82 &  & \phantom{1}9.6 ($\pm$1.1) &  & 8.98 ($\pm$0.5) \tabularnewline
 &  &  &  &  &  &  &  & \tabularnewline
$F_{12}$ &  & 39.9 &  & 39.3 &  & 37.0 ($\pm$7.9) &  & 37.0 ($\pm$5.6) \tabularnewline
 &  &  &  &  &  &  &  & \tabularnewline
\hline 
\end{tabular}
\end{table}

\end{widetext}

\begin{figure}[H]
\centering{}\includegraphics[scale=0.5]{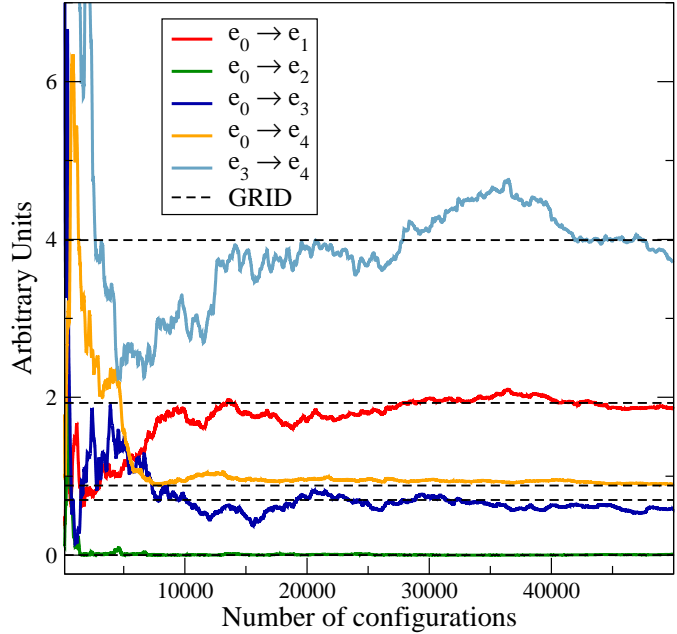} \caption{Running average of the Monte Carlo estimate for the oscillator strengths
of selected transitions for the isolated water molecule using Eq.
\ref{eq:trans-dip-integral-with-BM-MC}. Reference values, obtained
performing the integral on a numerical grid of $10^{6}$ points are
reported as dashed black lines.}
\label{fig:Fnm-MC-convergency} 
\end{figure}

The same oscillator strengths were also computed through the MC scheme
of Eq. \ref{eq:trans-dip-integral-with-BM-MC}. Results obtained upon
evaluation of the dipole over 25000 and 50000 structures are reported
in the last two columns of Tab. \ref{tab:h20-os-str-grid-MC}. The
errors in the MC simulations were calculated by evaluating the standard
deviation of the transition dipole moments. The resulting numerical
values are all consistent with the reference grid results within the
estimated error. Acceptable values have been obtained after 25000
MC configurations and are improved when considering 50000 MC configurations.
It is known that the number of points needed in a Monte Carlo simulation
grows polynomially with system dimensionality. This implies that calculations
of this kind can be performed with an affordable computational overhead
also for larger systems, where numerical integration on a grid becomes
overwhelming. It is worth stressing that the set of Monte Carlo points
generated, at which the molecular dipole has to be evaluated, is in
common for all transitions. However, convergence is not uniform: as
shown in Fig. \ref{fig:Fnm-MC-convergency}, it is slower for softer
vibrations (in this case, the bending) than for harder ones (bond
stretching). Moreover, it can be observed that, among stretching mode
transitions, the oscillator strength of the asymmetric stretch fundamental
excitation $(0\rightarrow4)$ converges faster than the symmetric
one $(0\rightarrow3)$. This is again due to the presence of the Fermi
resonance which affects states $e_{3}$ and $e_{2}.$ \textcolor{black}{Finally,
it is worth noting that knowledge of the oscillator strengths for
the selected transitions allows to determine the IR spectrum of water
at any temperature according to Eq. \ref{eq:abs-spec}.}

\section*{Summary and Conclusions}

In this work we introduced the possibility to calculate anharmonic
absorption intensities of vibrational spectra by means of semiclassical
dynamics. This is an important step in the direction of a complete
description of \textcolor{black}{infrared }spectroscopy with respect
to the power spectra simulations routinely provided by SC approaches.
The goal has been achieved by using harmonic vibrational states as
reference states to be evolved using the SC propagator (instead of
the commonly employed coherent states). In fact, by Fourier transforming
the SC recurring time-dependent overlap of harmonic states, the vibrational
eigenfunctions can be obtained through a decomposition on the harmonic
basis and the approach is readily extendable to high dimensional molecules.\textcolor{black}{{}
A successful application of the method, though, necessitates that
the eigenfunctions obtained within the harmonic approximation give
already a good qualitative representation. In fact we showed that,
in these cases, the most important anharmonic effects are already
included with only very few terms in the harmonic basis set.}

\textcolor{black}{In the spirit of multiple coherent semiclassical
dynamics we propagated just one trajectory per anharmonic state, so
that the n}umber of classical MD propagations is not directly related
to the size of the molecule but to the number of vibrational target
states. Once the trajectory associated to a target state has been
propagated, the semiclassical propagator is determined for all harmonic
basis functions needed to describe that specific state. The number
of basis functions to employ depends on the dimensionality of the
system, but it can be limited by employing an appropriate cutoff.

Test calculations performed on the 1D Morse oscillator have shown
that the harmonic basis is very well suited to describe the true anharmonic
vibrational states. In fact, all principal effects of anharmonicity
generally present in bond stretching vibrations are captured by including
only a few states in the harmonic expansion. Furthermore, excellent
accuracy of the eigenfunctions (compared with their analytical representation
used as a reference) is obtained after a simple post-processing refinement
consisting in a Gram-Schmidt orthonormalization.

The formalism was applied to the H\textsubscript{2}O molecule and
we demonstrated that also in this case very high quality eigenfunctions
(upon comparison to the DVR benchmarks) are obtained with a number
of expansion coefficients of the order of one dozen. The presence
of two main effects due to anharmonicity has been pointed out using
our method: i) the symmetric bond stretch actually provides an asymmetric
contribution that shifts the wavefunctions in the direction of bond
dissociation with respect to the harmonic counterparts, and ii) a
Fermi resonance arises between the fundamental of the symmetric stretch
and the overtone of the bending. The relevance of both effects has
been easily quantified on the basis of the harmonic expansion coefficients. 

Another advantage of the functional form of the harmonic basis is
that all the eigenfunctions are proportional to a multivariate Gaussian
function. We took advantage from this to compute the vibrational intensities
using a Monte Carlo strategy. First a number of the order of a few
thousand molecular configurations is generated at negligible computational
cost. Then, more expensive calculations are required to evaluate,
for each structure generated, the total molecular dipole (e.g. via
an \emph{ab initio} self consistent field calculation). This procedure,
given the good level of scalability of the MC integration, can be
directly extended to high dimensional systems with an affordable computational
cost. For the cases in which the application of this protocol is anyway
too demanding, we suggest (see Appendix B) a way to derive an approximate
estimate of the anharmonic oscillator strengths using a linear expansion
of the dipole operator. This approximation is commonly adopted to
derive oscillator strengths in the harmonic approximation and it just
requires calculation of the first order derivatives of the dipole
expectation value (a common output in most quantum chemistry packages). 

\section*{Acknowledgments}

Prof. Tucker Carrington is warmly thanked for reading the manuscript
and for interesting comments, and Dr. Gianluca Bertaina is thanked
for insights about the Monte Carlo application. The authors acknowledge
financial support from the European Research Council (ERC) under the
European Union\textquoteright s Horizon 2020 research and innovation
programme (grant agreement No {[}647107{]} -- SEMICOMPLEX -- ERC-2014-CoG).
We thank Università degli Studi di Milano for computational time at
CINECA (Italian Supercomputing Center), and CINECA under the Iscra-B
(grant QUASP) initiative for further computational time.

\section*{Appendix A\label{sec:APPENDIX-A}}

We derive analytically the result reported in Eq. \ref{eq:harm-coherent-overlap}
for the one dimensional scalar product $\braket{\phi_{k}|\tilde{Q},\tilde{p}}$
between the coherent state $\ket{\alpha}$ centered in $(\tilde{Q},\tilde{p})$
and the harmonic state of order k $\ket{\phi_{k}}$. This is easily
derived considering that coherent states are eigenstates of the harmonic
oscillator annihilation operator $\hat{a}=\sqrt{\frac{\omega}{2\hbar}}\left(\hat{Q}+\frac{i}{\omega}\hat{p}\right)$

\begin{align}
\hat{a}\ket{\alpha}=\alpha\ket{\alpha}\label{eq:ApA:alpha:def}
\end{align}
where

\begin{align}
\ket{\tilde{Q},\tilde{p}}=N_{\alpha}e^{i\eta_{\alpha}}\ket{\alpha}\label{eq:ApA:alpha:tilde}
\end{align}
with

\begin{align}
\alpha=\sqrt{\frac{\omega}{2\hbar}}~\tilde{Q}~+~i~\sqrt{\frac{1}{2\omega\hbar}}~\tilde{p}
\end{align}
In fact, by writing the coherent states in this convenient form, the
scalar product with the harmonic state is also straightforwardly derived
as:

\begin{align}
\braket{\phi_{k}|\alpha}=e^{i\eta_{\alpha}}e^{-\frac{|\alpha|^{2}}{2}}\frac{(\alpha)^{k}}{\sqrt{k!}},\label{eq:ApA:ska-wanted}
\end{align}
and the only term that remains unknown is the phase factor $\eta_{\alpha}$
needed to get to the coherent state definition in Eq. (\ref{eq:def-coherents}).
This is found comparing the scalar product between two coherent states

\begin{align}
\braket{Q_{1},p_{1}|Q_{2},p_{2}}=e^{i(\eta_{\alpha_{2}}-\eta_{\alpha_{1}})}\braket{\alpha_{1}|\alpha_{2}}\label{eq:ApA:co_st_superp1}
\end{align}
where $\alpha_{1}=\sqrt{\frac{\omega}{2\hbar}}Q_{1}+i\sqrt{\frac{1}{2\omega\hbar}}p_{1}$
and $\alpha_{2}=\sqrt{\frac{\omega}{2\hbar}}Q_{2}+i\sqrt{\frac{1}{2\omega\hbar}}p_{2}$.
The integral at the left hand side of Eq. (\ref{eq:ApA:co_st_superp1})
can be computed analytically giving

\begin{align}
\braket{Q_{1},p_{1}|Q_{2},p_{2}}= & e^{-\frac{\omega}{4\hbar}(Q_{1}-Q_{2})^{2}}\times\label{eq:ApA:co_st_superp_x1p1x2p2}\\
\times & e^{-\frac{1}{4\omega\hbar}(p_{1}-p_{2})^{2}}e^{\frac{i}{\hbar}(Q_{1}-Q_{2})(p_{1}+p_{2})}\nonumber 
\end{align}
while, the scalar product at the right hand side of Eq. (\ref{eq:ApA:co_st_superp1})
is:

\begin{align}
\braket{\alpha_{1}|\alpha_{2}} & =\sum_{k}\braket{\alpha_{1}|\phi_{k}}\braket{\phi_{k}|\alpha_{2}}=\label{eq:ApA:co_st_superp_alfabeta}\\
 & =e^{-|\alpha_{1}|^{2}}e^{-|\alpha_{2}|^{2}}e^{\alpha_{1}^{*}\alpha_{2}}\nonumber 
\end{align}
Inserting the results of Eq. (\ref{eq:ApA:co_st_superp_x1p1x2p2})
and Eq. (\ref{eq:ApA:co_st_superp_alfabeta}) into Eq. (\ref{eq:ApA:co_st_superp1}),
it is straightforward to obtain

\begin{flalign}
e^{i(\eta_{\alpha_{2}}-\eta_{\alpha_{1}})}=e^{\frac{i}{2\hbar}(Q_{1}p_{1}-Q_{2}p_{2})}\label{eq:ApA:phase_factor_alfa12}
\end{flalign}
and hence, for the generic scalar product in Eq. (\ref{eq:ApA:ska-wanted}),
it has to be

\begin{flalign}
\eta_{\alpha}~=~-\frac{1}{2\hbar}\tilde{Q\,}\tilde{p}\label{eq:ApA:phase_factor_alfa_xp}
\end{flalign}

\section*{Appendix B\label{sec:Appendix-B}}

In view of the application of our methodology to the calculation of
the oscillator strengths for high dimensional molecular systems, where
the Monte Carlo sampling can become computationally demanding, it
is worth noting that the integral in Eq. (\ref{eq:trans-dip-integral})
can be evaluated in an approximate way without any further sampling
of the molecular configurational space. This is achieved by considering
the common linear expansion of the molecular dipole:

\begin{align}
\mu_{0N}(\mathbf{q}) & -\mu_{0N}(\mathbf{q}_{eq})\simeq\nonumber \\
\simeq & \sum_{\alpha=1}^{N_{v}}\left.\frac{\partial\mu_{0N}}{\partial q_{\alpha}}\right|_{\mathbf{q}_{eq}}(q_{\alpha}-q_{eq,\alpha})=\mathbf{Z}_{q}\cdot\mathbf{Q},\label{eq:DH-approx}
\end{align}
where $\mathbf{Z}_{q}=\left.\frac{\partial\mu_{0N}}{\partial q_{\alpha}}\right|_{\mathbf{q}_{eq}}$
. Using this linearization, the transition dipoles can be obtained
directly from the expansion of the eigenstates in the harmonic basis
derived with the SC calculation

\begin{align}
\bra{e_{n}}\hat{Q}_{\alpha}\ket{e_{m}}=\label{eq:dip-on-h-basis}\\
=\sum_{\mathbf{K},\mathbf{K}'}C_{n,\mathbf{K}}C_{m,\mathbf{K}'} & \bra{\mathbf{K}}\hat{Q}_{\alpha}\ket{\mathbf{K}'}\nonumber 
\end{align}

\noindent and by computing the transition dipoles on the harmonic
states using the following relations:

\begin{align}
 & \bra{\mathbf{K}}\hat{Q}_{\alpha}\ket{\mathbf{K}'}=\left(\ \prod_{\beta\ne\alpha}^{N_{v}}\delta_{K_{\beta},K'_{\beta}}\right)\sqrt{\frac{1}{2\omega_{\alpha}}}\times\nonumber \\
 & \left(\delta_{K_{\alpha}+1,K'_{\alpha}}\sqrt{K_{\alpha}+1}+\delta_{K_{\alpha}-1,K'_{\alpha}}\sqrt{K_{\alpha}}\right)
\end{align}

\noindent that are immediately derived using the fact that $\hat{Q}_{\alpha}=\sqrt{\frac{1}{2\omega_{\alpha}}}\left(\hat{a}_{\alpha}^{\dagger}+\hat{a}_{\alpha}\right)$,
with $\hat{a}_{\alpha}^{\dagger}$ and $\hat{a}_{\alpha}$, respectively,
the harmonic oscillator creation and annihilation operators for normal
mode $\alpha$.

\bibliographystyle{aipnum4-1}
\bibliography{SCARRAFONE}

\end{document}